# Intrinsic-Stabilization Uni-Directional Quantum Key Distribution Between Beijing and Tianjin

Xiao-fan Mo[1], Bing Zhu[1, 2], Zheng-fu Han[1*], You-zhen Gui[1], Guang-can Guo[1]

[1]Key Lab of Quantum Information

[2]Department of Electronic Engineering and Information Science

University of Science and Technology of China

Hefei, Anhui, 230026, P. R. China

**Quantum key distribution provides unconditional security for communication. Unfortunately, current experiment schemes are not suitable for long-distance fiber transmission because of instability or backscattering. We present a uni-directional intrinsic-stabilization scheme that is based on Michelson-Faraday interferometers, in which reflectors are replaced with 90° Faraday mirrors. With the scheme, key exchange from Beijing to Tianjin over 125 kilometers with an average error rate is below 6% has been achieved and its limited distance exceeds 150 kilometers. Experimental result shows the system is insensitive to environment and can run over day and night without any break even in the noise workshop.**

For the security of sensitive data transmission, a key must be securely exchanged firstly between two users. Quantum cryptography is the only way to distribute the key with unconditional secrecy through the public channel [1,2]. Up to now. Previous experimental demonstrations have been implemented over fiber [3-6] and free space [7-9]. In the near future, quantum channel is mostly commercial communication fiber cable, which is normally immerged in a complex and uncontrollable environment. Therefore, a practicable QKD scheme should be insensitive to the environment and be immune from the back scattering [10,11], especially for long distance quantum transmission. The double unbalance M-Z interference scheme presented by Bennett [12] lets the two photon pulses go along the same routes, this ingenious design improves the stability of QKD, but the instability is obstinate [10,13,14]. The longer the quantum channel is, the faster fluctuating of the interference visibility will be [15]. Two groups presented a perfect stability QKD scheme with a round-trip configuration almost at the same time [14,16]. However, this scheme suffers from the backscattering and is hard to withstand the Trojan-horse attack [11]. So a uni-directional scheme becomes necessary for long distance quantum key distribution [4].

Here, we report a Michelson-Faraday scheme in which the encoder and decoder are equal at all. Each coder is a modified unbalance fiber Michelson interferometer, shown as Fig.1. A coder is made of a 3 dB conventional fiber (Corning SMF-28) coupler C which splits any input photon pulse into two ones, a piece of 1.5 meters same type fiber which brings about 7.5ns time delay between two photon pulses, an integrated optical phase modulator (LiNbO$_3$ wave-guide) PM inserted in one arm of the interferometer for phase coding, and two 90° Faraday mirrors FM located at end of each arm respectively for reflecting the photon pulse back and turning it's polarization for 90° at the same time. So, each pulse will run a round trip along its arm with orthogonal polarization. It is easy to certify that the JONES matrix of each arm is unitary, even if the two arms are unequal in length and different on polarization dependence. This just meets the anti-disturbance conditions equation (7) in Ref. [15], namely this scheme is free of any polarization dependence disturbance not only from the long distance quantum transmission channel but also from inside of the encoder and decoder. So we named the scheme intrinsic-stabilization quantum key distribution system. Based on this scheme, the interference visibility of the system with the quantum channel as long as 175 km, is measured and shown at Fig. 2a. It shows us a long period stability even though the system is operated in a noise workshop without any special vibration isolation or damping. It is necessary to point out that there is still a residual phase drift in the system (see Fig. 2b), as we know, this can be mainly own to the temperature fluctuation and also a little bit contribution from the vibration of the two coders. But this phase drift is very slowly and thus easy to be corrected timely. As an accessory, an optical circulator Cir and a photon detector D are arranged at the output of Alice's security zone and the input of Bob's security zone. On Alice's side, they are used to inspect the possible Trojan-horse photons, while they are used not only for security monitor but also for separating and detecting the coding photons on Bob's side.

In order to prove the design, an experimental system is measured in lab, with an ID200 photon detector (Switzerland, University of Geneva), whose measured minimum dark counts rate is $8\times 10^{-7}$/pluse. A strongly attenuated 1550nm laser pulse imitates the single photon source. For the secrecy point's of view, the coded pulse is fixed as 0.1 photons/pulse with 1 ns width at the output of Alice's security zone, but the uncoded one is about 0.4 photon/pulse because of the additional attenuation of the integrated phase modulator. Fortunately this multi-photon pulse is harmless to the security because it carries not any information. To conquer the instability of the working point because of the remnants of the phase drift, the periodic phase testing and rectifying are adapted. In fact, the processes of quantum key distribution are operated as: testing and correcting the phase firstly, carrying out QKD secondly, and repeating the above process. On average, the time for phase testing and rectifying is less than ten percent of the total operation time. If make the working frequency of the system risen, this rate will become lower. Fig. 3 shows us the relationship of the experimental transmission distance and the error rate. When the error rate 10% is considered as secure limit, we have realized a long-term stable quantum key distribution over 150 km in lab.

For the practical quantum key distribution between the cities, we select one station in Beijing and the other one in Tianjin (fig. 4). Two stations are connected with communication fiber cable of China Network Communications Group Corporation (CNC); the total fiber length is more than 125 km, with the total attenuation 26 dB. The result shows that our QKD system can run over day and night without any break in the noise workshop, the long-term error rate is below 6% which is nearly 1% higher than that in lab at the same transmission distance.

In conclusion, we have designed a simple and intrinsic-stabilization uni-directional quantum key distribution scheme. This schematic configuration can resist to all the disturbances from the long distance transmission channel and all the disturbances of polarization dependent imposed on Alice's and Bob's devices. Basing on it, we have realized a long-term stable quantum key distribution over 150 km in lab and performed a key exchange over 125 km between Beijing and Tianjin, two cities of China. This means that the true-life QKD communication is coming now.

This work was funded by the National Fundamental Research Program of China (Grant No. 2001CB309301), also by the National Natural Science Foundation of China (Grant No. 60121503) and the Innovation Funds of the Chinese Academy of Sciences.

* **Correspondence** and requests for materials should be addressed to Z. F. Han (zfhan@ustc.edu.cn).

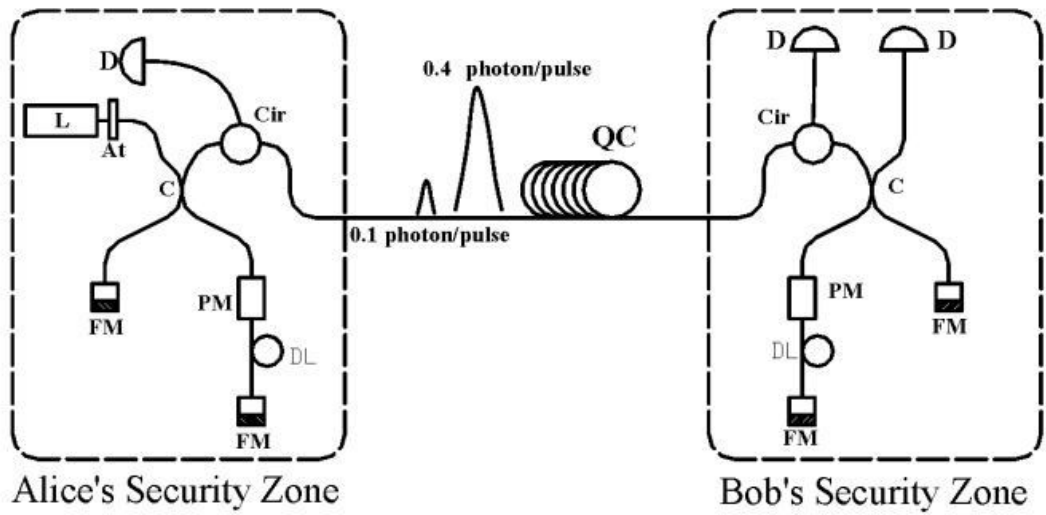

Fig 1. Intrinsic-Stabilization Uni-Directional Quantum Key Distribution

L— Faint Pulse Laser, At—Optical Attenuator, D—Single Photon Detector, Cir—Optical Circulator, C—3dB Fiber Coupler, PM—Phase Modulator, DL—Delay Line, FM—90° Faraday Mirror.

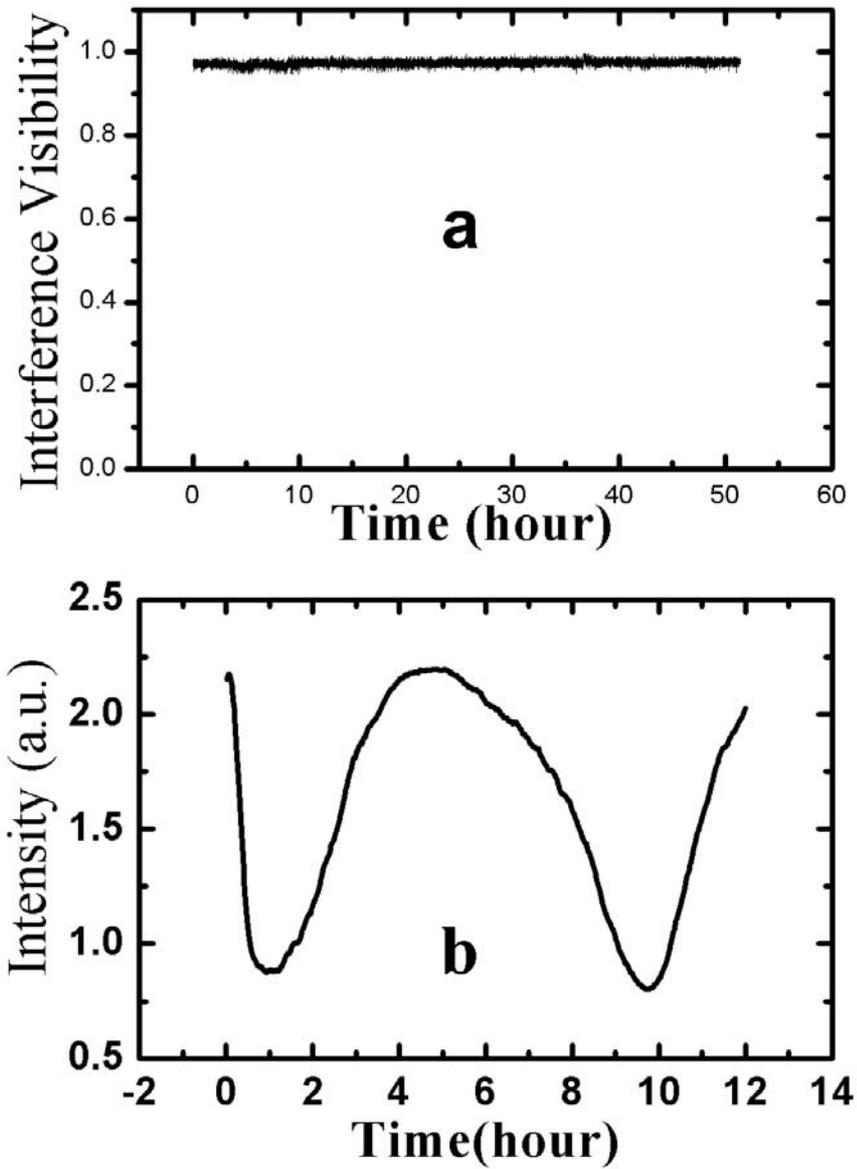

Fig 2. Long-term stability of Michelson-Faraday QKD system

a) Interference visibility with 175km fiber as quantum channel, the encoder is modulated by periodic voltage for test. b) Phase drift of the system without modulating signal.

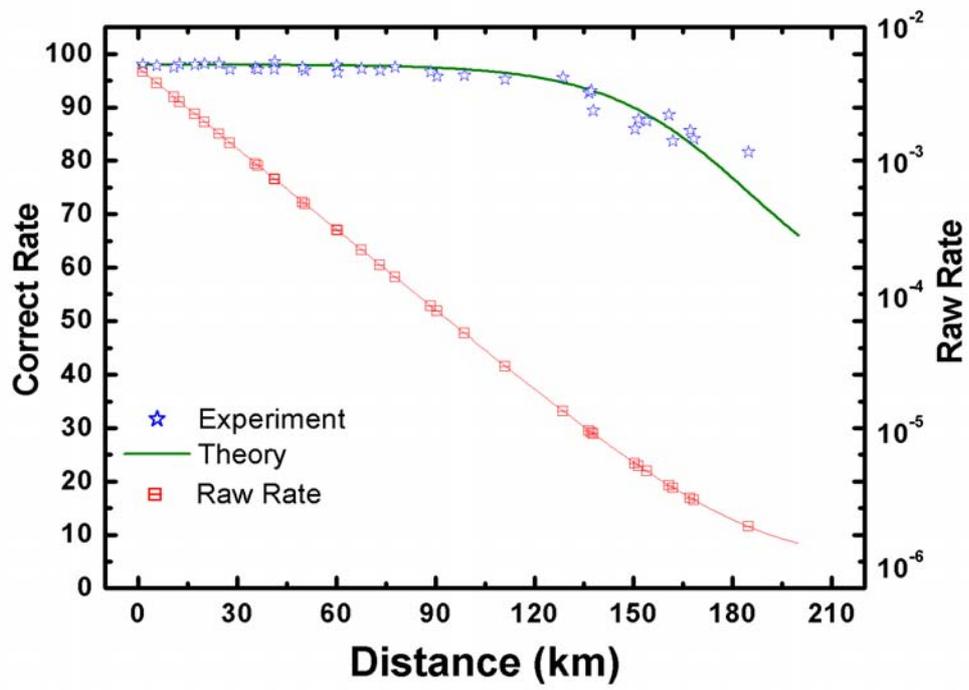

Fig 3. Correct bit rate and the raw bit rate to the transmission distance in Lab

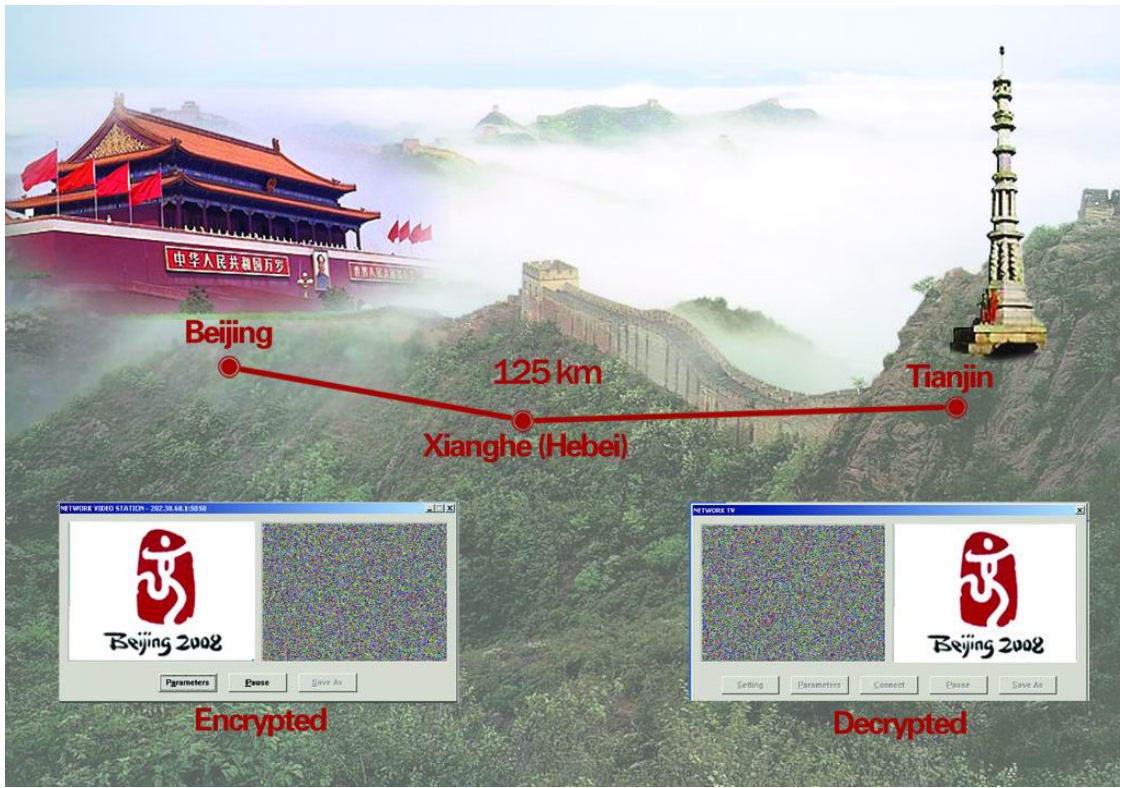

Fig 4. Quantum Key Distribution from Beijing to Tianjin city over 125km, the inserts are original encrypt and decrypt pictures on computers of Alice's and Bob's side respectively.